\documentclass [preprint,showpacs,prl]{revtex4}
\usepackage{bm}
\begin{document}
\title{Metal nanoplasmas as bright sources of hard x-ray pulses}   
\author{P. P. Rajeev, P. Taneja, P. Ayyub, A. S. Sandhu, and G. R. Kumar\footnote{electronic address: grk@tifr.res.in}}
\address{Tata Institute of Fundamental Research, 1, Homi Bhabha 
Road, Mumbai 400 005, India.}
\date{\today}
\newcommand{\e}{\mbox{\boldmath$\eta$}}
\newcommand{\x}{\mbox{\boldmath$x$}}
\newcommand{\si}{\mbox{\boldmath$\xi$}}

\begin{abstract}
	We demonstrate a 13-fold increase in hard x-ray bremsstrahlung yield (10 - 200 keV) emitted by a copper plasma created by 100 fs, 806 nm pulses at $10^{14}-10^{15}$ Wcm$^{-2}$. This enhancement is achieved by depositing a thin film of  copper nanoparticles of size ~15 nm, on the target surface. A simple model that invokes local field modifications by surface plasmon excitation and `lightning rod' effects explains the observed enhancement quantitatively and provides pointers to the design of structured surfaces for maximizing the emission. 

\end{abstract}
\maketitle

	The interaction of intense, ultrashort laser pulses with solid plasmas is of immense interest not only from a basic physics point of view but also due to their promise as micron-sized sources of ultrashort x-rays, in areas like lithography and time resolved diffraction \cite{Jiang,Siders}. Methods to enhance the x-ray yield are of great importance, and the influence of various laser and target conditions is widely investigated. Pre-formed plasmas \cite{Stearns,Nishikawa} yield significant enhancements at the cost of an increase in the x-ray pulse duration.  Recent literature reports impressive enhancements in soft \cite{Murnane} and moderately hard x-ray regions \cite{Wulker} using structured surfaces, viz. gratings \cite{Murnane,Gauthier,Gordon}, ``velvet" coatings \cite{Kulcsar}, porous and nanocylinder \cite{Nishikawa,Gordon,Nishikawa2} targets. However, little attention has been paid to examine ways of improving the very hard ($>$~10~keV) x-ray yield, a signature of hot electrons created in the plasma mainly by resonance absorption (RA) \cite{Kruer}, in the intensity regime $10^{14}-10^{15}$ Wcm$^{-2}$. Enhanced x-ray yield, therefore, implies enhanced hot electron production, an issue that is of immense interest to inertial fusion research \cite {Tabak} and particle acceleration \cite{Wilks}. 

	In this Letter, we demonstrate a method to enhance hard x-ray bremsstrahlung by coating nanoparticles (NP) on optically smooth metal targets. We report a 13-fold enhancement in the total x-ray emission in the 10-200 keV range using copper NP coated targets and show that they yield hotter electrons in comparison to optically polished, uncoated targets. These results are well explained by a simple model that invokes local electric field enhancement via surface plasmon excitation and `lightning rod' effects. In addition, our model  provides  clear guidelines to the design of  surface microstructures that would maximize hot electron and x-ray yields. The giant enhancement in nonlinearity that such fractal structures provide \cite{Shalaev} and the ease of patterning such targets can prove invaluable in the design of efficient ultrashort x-ray sources.

	Our Ti:Sapphire laser system generates 806 nm, 100 fs (FWHM) pulses with 5 mJ energy per pulse. The pulses have a contrast ratio of $10^5$ with the pedestal in picosecond timescales. A detailed description of the experimental set-up can be found elsewhere \cite{Sudeep}.  P-polarized laser pulses are focused on targets housed in a vacuum chamber at $10^{-3}$ Torr. A 1mJ laser pulse yields a peak intensity of $1.3\times10^{15}$ Wcm$^{-2}$ at a typical focal spot size of 40 $\mu$m. The target is constantly rotated and translated in order to avoid multiple laser hits at the same spot. X-ray emission from the plasma is observed in the plane of incidence, at $45^{\circ}$ to the  target normal, with a NaI (Tl) scintillation  detector, wherein energy of the incident photon determines the amplitude of the PMT signal yielding energy-dispersed yields in the observed energy range. The detector is gated with the laser pulse and the signal is collected only in a time window of 30 $\mu$s, to ensure background-free acquisition. The observed spectrum is essentially bremsstrahlung as the characteristic emissions are blocked by the 5-mm BK-7 window of the chamber.

	We use two types of Cu NP coated targets, one with spherical (inset of Fig. ~1) and the other with ellipsoidal nanoparticles (SNP and ENP, respectively). Their emission is compared with highly polished copper targets. Cu nanoparticles are deposited by high pressure dc sputtering  \cite{Praveen} on such polished copper discs (held at $0^{\circ}$C for spherical particles and at $-50^{\circ}$C for ellipsoids). The resulting nanocrystalline Cu films are  optically flat and 1 $\mu$m thick. The coherently diffracting crystallographic domain size ($d_{XRD}$) is obtained from x-ray diffraction line broadening, using the Scherrer technique. For a film deposited in 180 mTorr Ar environment at a sputtering power of 200W, we obtain $d_{XRD}$ = 15nm. The aspect ratio is obtained from a comparison of $d_{XRD}$ calculated from (111) and (200) diffraction lines. For basic optical characterization, the linear absorption spectra are measured (Fig.~1). The ENP (aspect ratio $\sim$ 1.5) have comparatively less absorption in the linear regime, in contrast to their behavior at higher intensities, as discussed later.  
	
	 The reflectivity data were fitted to the Drude model for $\lambda > 650$ nm, below which the inter-band transitions contribute to the dielectric function. Assuming a constant plasma frequency and with a collision frequency adjusted for a best fit, we get $\epsilon$ as a function of $\lambda$. The fits yield effective permittivity ($\epsilon'+i\epsilon''$) of the NP-void composite. Using the generalized Bruggeman effective medium approximation \cite{Granqvist}, the permittivities of the SNP and ENP are obtained as $-27 + i33$  and $-27 + i44.4$ respectively, as opposed to their bulk value of $-27 +i2.5$. Extensive studies have been carried out on the variation of the dielectric constant with particle size \cite{Kreibig,Exp}. The real part of the dielectric constant is shown to be unaffected in most systems, unless the particle size is extremely small. The imaginary part increases due to the limited electron mean-free path in the NP  \cite{Kreibig2}. The imaginary parts mentioned above are, however, much larger than the theoretically predicted values  \cite {Kreibig2} and this mismatch is likely to be due to the factors like particle size distribution and dipole interactions between particles that are excluded in the modeling. However, this discrepancy does not significantly affect the interpretations of our results, as will be evident later.
 
	To elucidate the role of nanoparticles in hot electron generation in two distinct regimes of RA, we present hard x-ray bremsstrahlung data at two angles of incidence, $10^{\circ}$ and $45^{\circ}$.  For a flat target, collisional absorption is expected to be the main contributor at $10^{\circ}$, while RA is expected to play a major role at $45^{\circ}$.  Fig.~2 presents a comparison of bremsstrahlung emission for $10^{\circ}$, measured at a solid angle of 720 $\mu$Sr, from the polished and SNP-coated copper targets irradiated at $2.0\times10^{15}$ Wcm$^{-2}$. The total energy emitted per laser pulse in the above range from the polished target is $2.2\times10^{-12}$ J, whereas it is $1\times10^{-11}$ J from the nanoparticle coated target, assuming isotropic emission. It is clearly evident that there are two temperature components (6 keV and 14 keV) in the spectrum from  the NP-coated target, while the higher component is insignificant in the emission from the polished target.

	  The hot electron temperature given by RA is expected to follow the scaling law \cite{Zhang} $T_{hot}=6\times10^{-5}(I\lambda^2(W cm^{-2} \mu m^2))~^{0.33}$, where  $I$ is the intensity and $\lambda$ is the laser wavelength. These parameters, at $2\times10^{15}$ Wcm$^{-2}$ yield  $T_{hot}$ = 5.8 keV, close to the observed hot electron temperature for polished targets. The NP coated targets, however, yield a significantly higher temperature component, which arises from the excess absorption caused by local field enhancements as  we show below.

	 Modification of electric field due to surface protrusions has been well studied in connection with  second harmonic generation \cite{Shen} and surface enhanced Raman scattering \cite{Gersten}.  At higher intensities,  the electric field resonance is known to be a major source of hot electrons in a cluster plasma \cite{Ditmire}. However, this idea has not been utilized so far to understand the excess absorption of intense laser light on modulated surfaces.

	For simplicity, the NP target is modeled as a collection of hemispheroids of permittivity $\epsilon$, embedded on a flat substrate kept in vacuum, as shown in the inset of Fig.3 (a). Consider a p-polarized wavefront of amplitude $E$, incident at an angle $\theta$ to the major axis of the spheroid. The model becomes much simpler with the assumption that the field along the major axis alone contributes to the enhancement. Thus, the resultant electric field at any point on the surface of the hemispheroid becomes 

\begin{equation}
{\bf E_r} = {\bf E_L^{surf}} + E \cos{\theta}\hat{{\x}},
\end{equation}

where ${\bf E_L^{surf}}$ is the locally enhanced field and $E\cos{\theta}$ is the tangential component of the incident electric field on the metal surface. The enhanced local field on the surface of the spheroid can be computed as \cite{Shen}

\begin{equation}
{\bf E_L^{surf}} =  [ L_\parallel^{surf}\sin\alpha\hat{{\e}} + L_\perp^{surf}\cos\alpha\hat{{\si}} ]E \sin{\theta},
\end{equation}

where $L_\parallel^{surf}$ and $L_\perp^{surf}$ are  the local field correction factors given by

\begin{equation}
L_\perp^{surf} = L_R\epsilon/ [\epsilon - 1 + L_R [1 + i\frac{4\pi^2V}{3\lambda^3}(1-\epsilon)]]
\end{equation}

and $L_\parallel^{surf} =  L_\perp^{surf}/\epsilon$, which is absent for metals, as there cannot be a parallel component of electric field on the metal surface. $L_R$ is the `lightning rod' factor defined as $L_R ~=~ 1 ~- ~\xi Q^ \prime(\xi)/Q(\xi)$, where $\xi~=~ [1 - (b/a)^2]^{-1/2}$ and $Q(\xi) = (\xi/2)ln [(\xi+1)/(\xi-1)] - 1$. $V$ is the volume of the spheroid. The maximum enhancement occurs towards the tip of the structure (low $\alpha$ values). The {\it effective intensity} at the tip ($\alpha = 0^{\circ}$) can be written as

\begin{equation}
I_{r} = I_{in} [(L_\perp^{surf})^2 \sin^2{\theta} + \cos^2{\theta}].
\end{equation}

	The local field correction factors have a resonant behavior with $a/b$, the aspect ratio of the spheroid, as shown in fig.3 (a). The dielectric constant, $\epsilon$, is assumed to be a function of the particle diameter, viz. $\epsilon_{nano} = \epsilon'_{bulk} + i\epsilon''_{bulk}(1 + l/b)$, where $l$ is the mean free path of the electrons \cite{Kreibig}. As is evident, silver NPs with aspect ratios close to the resonance values will yield greater enhancement than  similar ones of gold and copper, although the absolute values could be affected by the possible plasma screening effects for large aspect ratios.   Fig. 3 (b) provides the variation of dielectric functions of bulk Cu and ENP with input laser intensity, derived from self-reflectivity measurements \cite{Downer}. The deviation from room temperature values occurs only above $10^{14}$ Wcm$^{-2}$. In fig. 3 (a), the resonance behavior depends on $\epsilon$, although $L_\perp^{surf}$ does not depend critically on $\epsilon$ for small $a/b$ values. Thus, in the present study, the room temperature values of $\epsilon$ will be quite good approximations in the calculations as $L_\perp^{surf}$ will be more or less the same even for a drastically different plasma $\epsilon$ (`stars', fig.3 (a)) and the plasma shape remains intact as it is a femtosecond interaction. 
 
	Substituting the value of  $L_\perp^{surf}$ for the spherical particle, we obtain $I_{r}/I_{in} \sim 1.4$, at $\theta = 10^\circ$. Thus, \emph{the NP-coated target is equivalent to a polished target with a rescaled (enhanced) intensity}. Fig.2 (inset) provides a comparison of the original data of yields of SNP and polished targets with  the data obtained by rescaling the points for the NP target by $I_{r} = 1.4I$, and they are in reasonably good agreement. The apparent convergence of the yields was due to the peeling of the NP coating at high intensities and does not indicate any closure of the nanostructures.

	Fig.~4 presents a comparison of bremsstrahlung emission, measured at a solid angle of 22 mSr, from the polished, spherical and ellipsoidal NP-coated Cu targets  irradiated at $45^{\circ}$, at $6.0\times10^{14}$ Wcm$^{-2}$. The total energy emitted per pulse from a polished target is $4.2\times10^{-14}$ J, while it is $5.7\times10^{-13}$ J using the ENP target, assuming isotropic emission.  This amounts to a 13-fold enhancement in hot electron production at an intensity which is less than half of that used for $10^{\circ}$ incidence. The spherical nanoparticles yield $1.4\times10^{-13}$ J, giving $\sim$ 3-fold enhancement as at the lowest intensity at $10^{\circ}$. Two temperature components, 3 keV and 11 keV, are observed in the spectrum from both NP-coated targets, whereas the higher component is hardly present in the emission from polished target. That the ellipsoidal nanoparticles give more than 4 times yield than the spherical particles is easily understood from Fig.~3 to be due to the enhancement of both `lightning rod' effect and plasmon resonance.  An intensity rescaling, as above, gives $I_{r}/I_{in} \sim 9$, almost in agreement with the x-ray yield enhancement as predicted by RA ($\sim I^{4/3}$).  The RA scalings are equally applicable for NP targets as their dimensions are much smaller than $\lambda$. The higher component $T_{hot}$ obtained using ENP target corresponds to that obtained using a polished target at $9\times10^{15}$~W~cm$^{-2}$ (close to the rescaled intensity for the ellipsoidal particle), as reported in our earlier work \cite{Rajeev}.

	The field enhancements are also responsible for enhanced nonlinearities \cite{Shalaev}.  Thus, multiphoton ionization in an NP target will be significantly enhanced as compared to the polished targets, which, in turn results in denser plasma formation and excess energy absorption. Due to the enhanced local fields and finite size of the surface protrusions, the oscillation energy of electrons reach a high value even at moderate intensities \cite{Kupersztych}.   At higher intensities, these `hot' electrons yield bremsstrahlung and characteristic emissions in hard x-ray regime, when they undergo collisions.

	Intensity enhancements by the surface structures could form pre-plasma for low contrast pulses, which would destroy the structure and adversely affect enhancement by the main pulse \cite{Kulcsar}. In our case, however, pre-plasma formation is negligible as the threshold intensity for plasma formation is $10^{12} - 10^{13}$ Wcm$^{-2}$ \cite{YZhang}, and our pre-pulses have  intensities lower than this, given our contrast and enhancement factors. The observed enhancements themselves substantiate the integrity of the structure and the irrelevance of pre-formed plasma in our case.  
	
	In summary, we report a 13-fold enhancement of the total bremsstrahlung emission in the 10-200 keV range from a copper nanoparticle-coated target in comparison to an optically polished Cu target. A simple model of the surface shows that enhancements in local electromagnetic fields result in excess absorption and hotter electrons that, in turn, enhance x-ray emission. Such enhanced emission is very attractive for practically viable hard x-ray sources. Further, our model provides guidance for designing better sources of ultrashort hard x-ray pulses; further studies are in progress towards this goal.  The intensity levels we have used in this study are quite modest and are easily available from modern femtosecond lasers operating at multi-kHz repetition rates \cite{Murnane2}, making applications of nanoparticle coated targets very promising.

	The authors thank N. Kulkarni for help with experiments and M. Krishnamurthy, V. Kumarappan and D. Mathur for useful discussions. The high energy, femtosecond laser facility has received substantial funding from the Department of Science and Technology.

\newpage

\begin{figure}
\caption{Linear absorption spectra of polished and nanoparticle coated targets. Inset: SEM image of spherical NP coated target. The solid lines show the Drude model fits.}
\end{figure}

\begin{figure}
\caption{Bremsstrahlung emission from smooth and SNP coated targets at $10^{\circ}$. The inset shows the variation of integrated emission with input laser intensity.}
\end{figure}

\begin{figure}
\caption{(a):Enhancement factor at $\lambda$ = 806 nm as a function of the aspect ratio for different metals, for b = 7nm. $L_\perp^{surf}$ for $\epsilon = 10 + i30$ is shown stars. Inset: spheroidal model for Nanoparticle. (b): Variation of dielectric functions of bulk Cu and ENP with intensity}
\end{figure}

\begin{figure}
\caption{Comparison of bremsstrahlung emission from polished and spherical and ellipsoidal NP coated targets at $45^{\circ}$. The exponential fits yield temperatures.}
\end{figure}
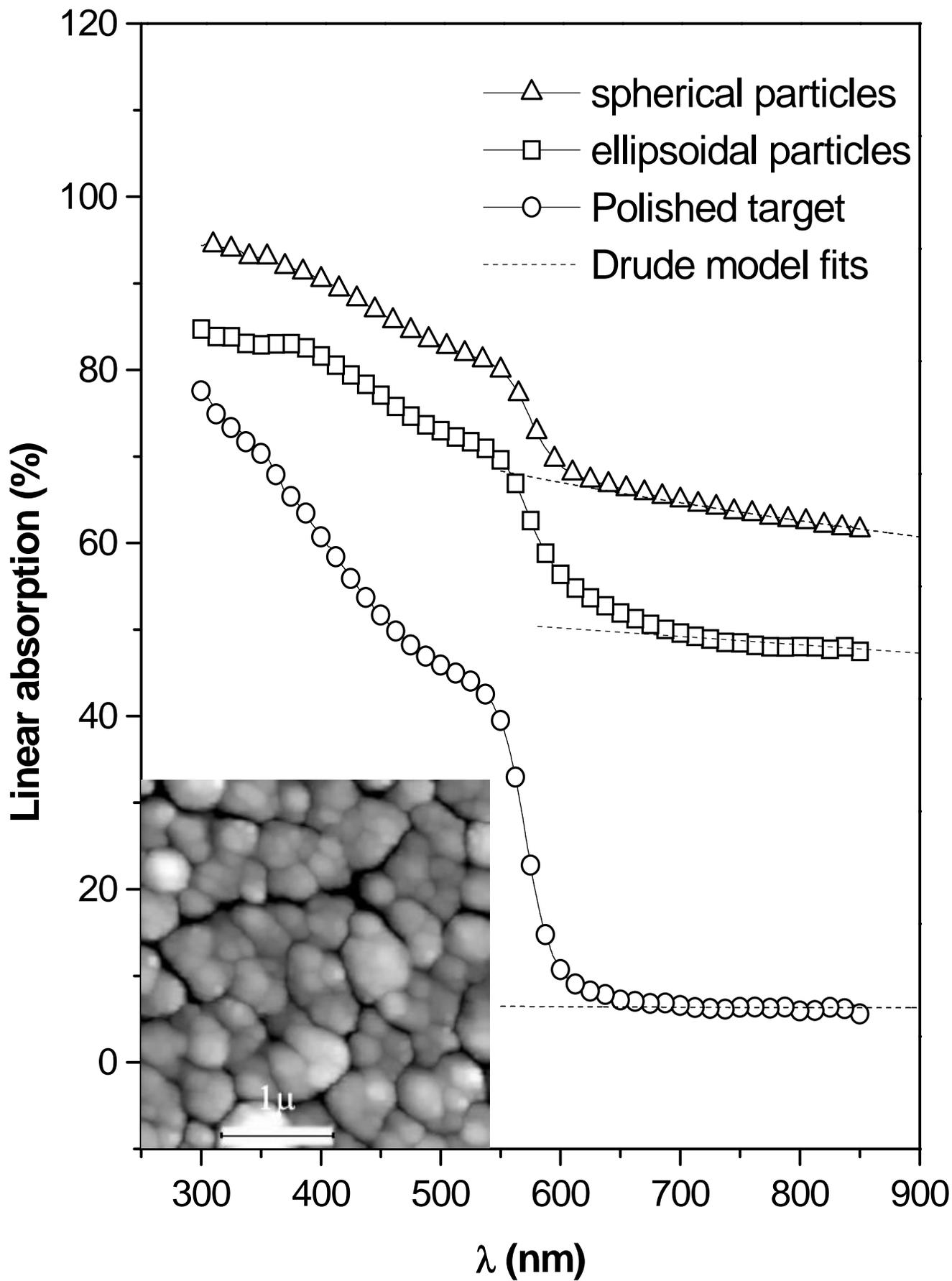
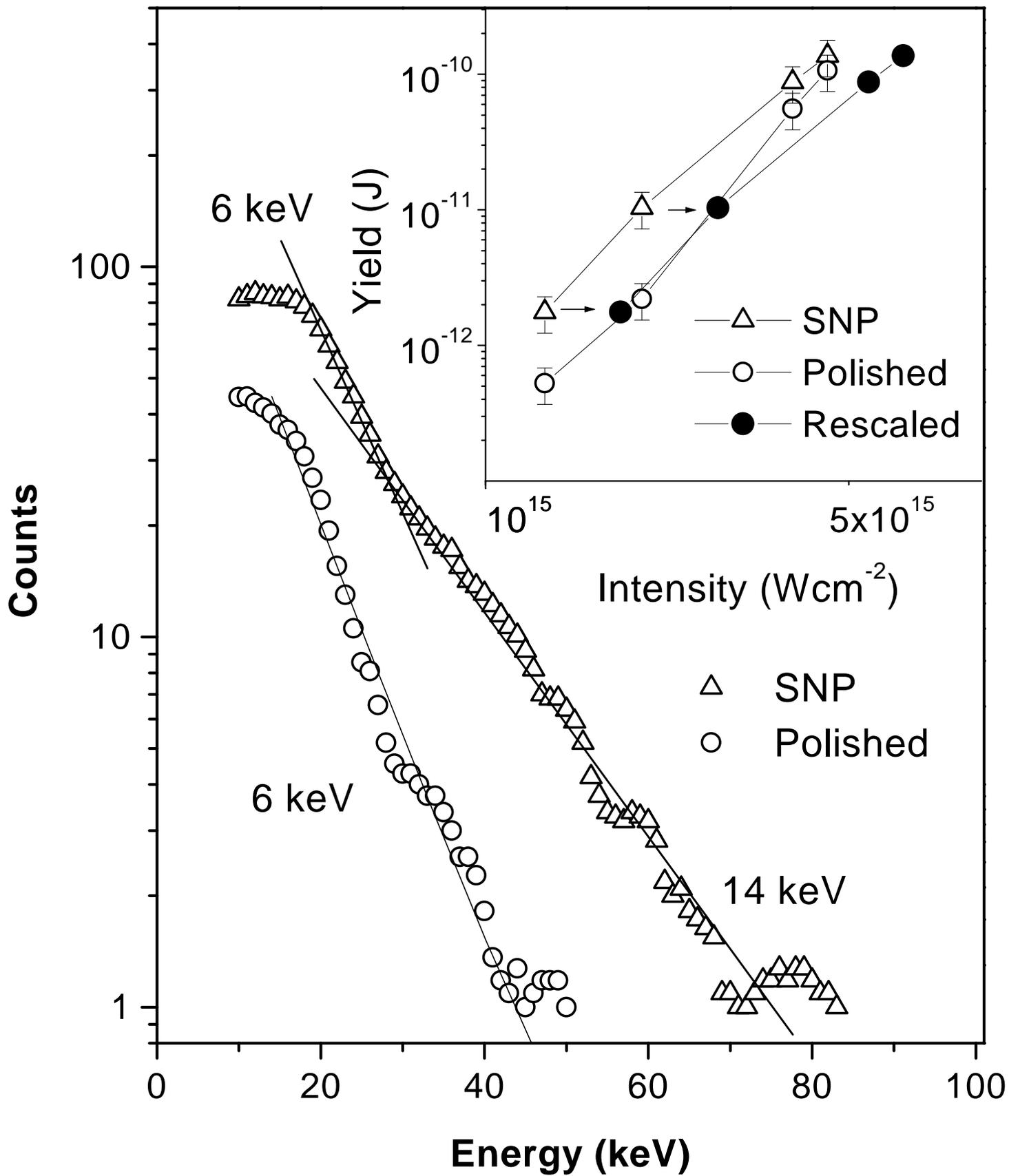
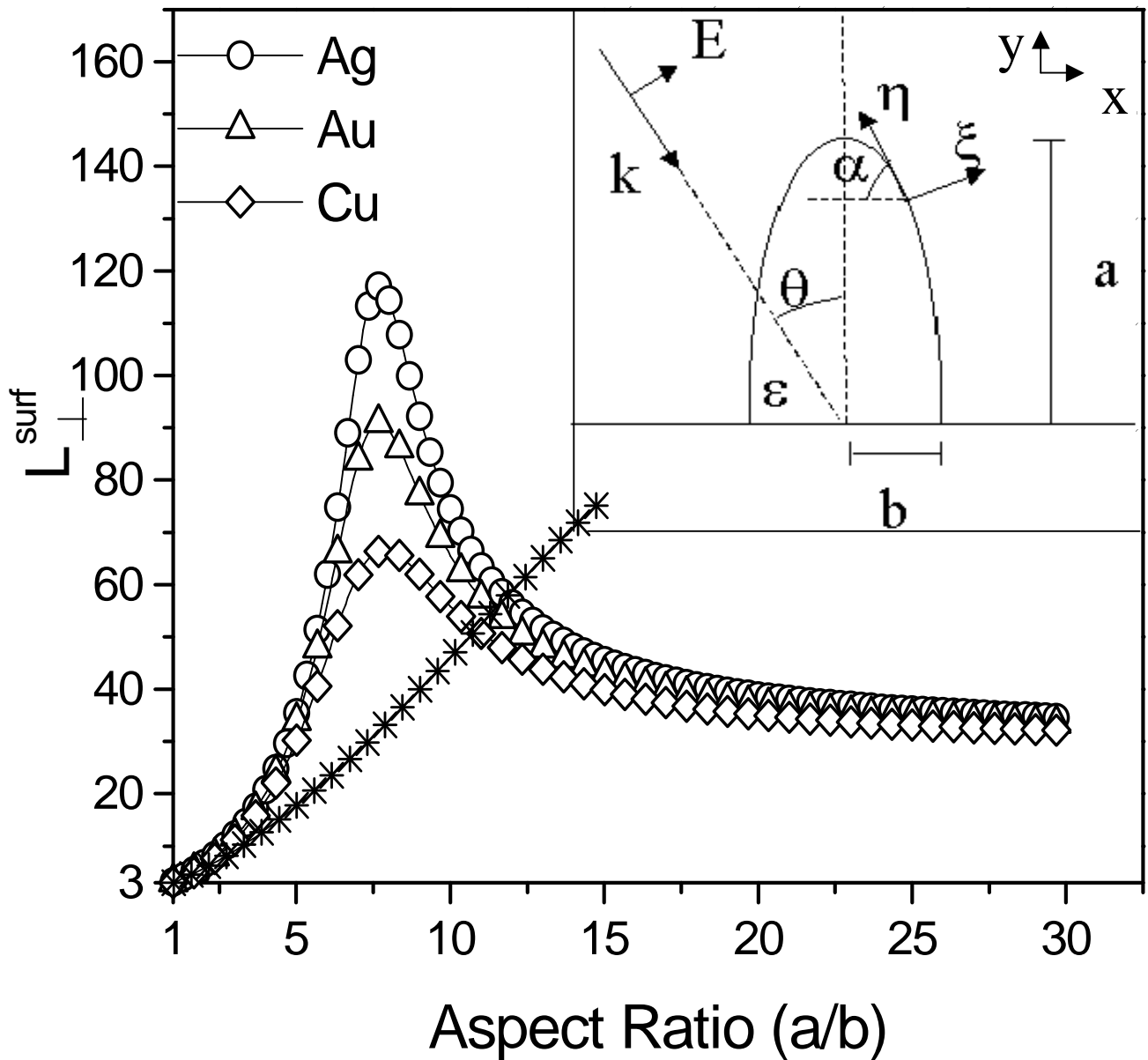
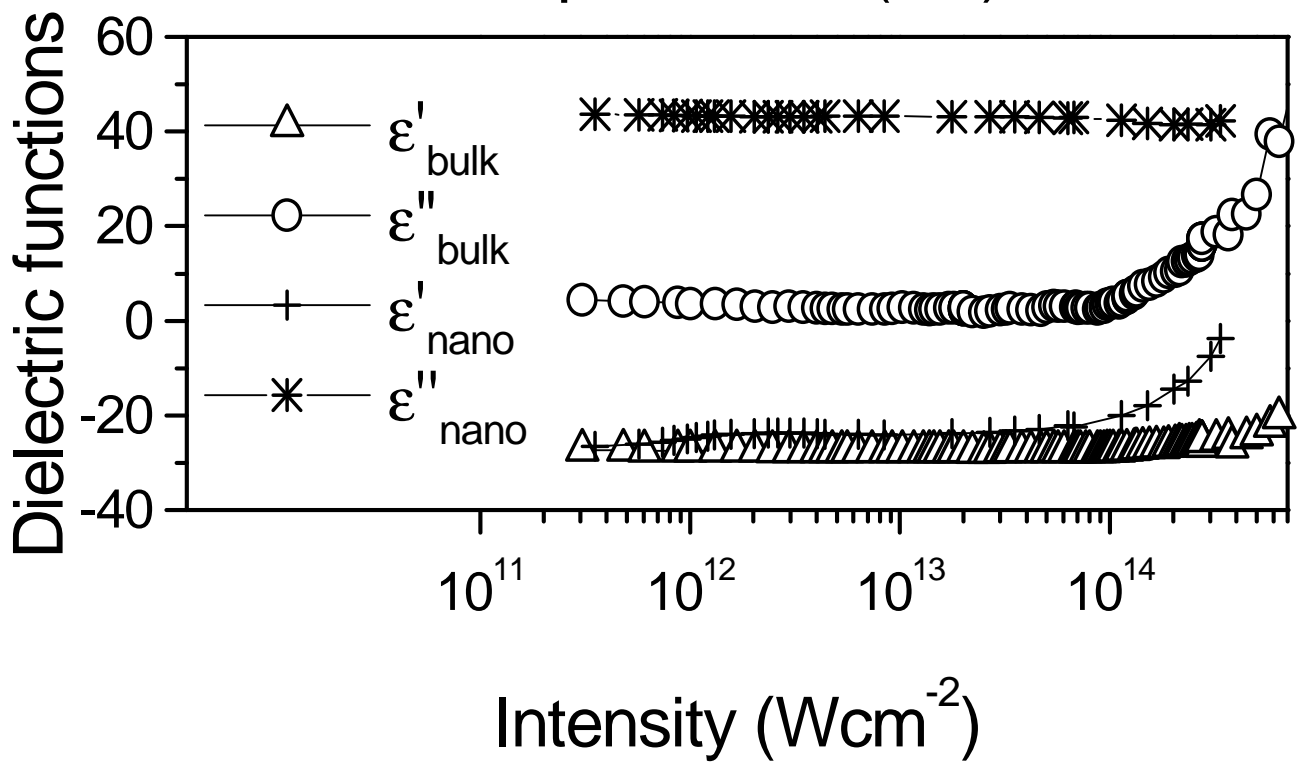
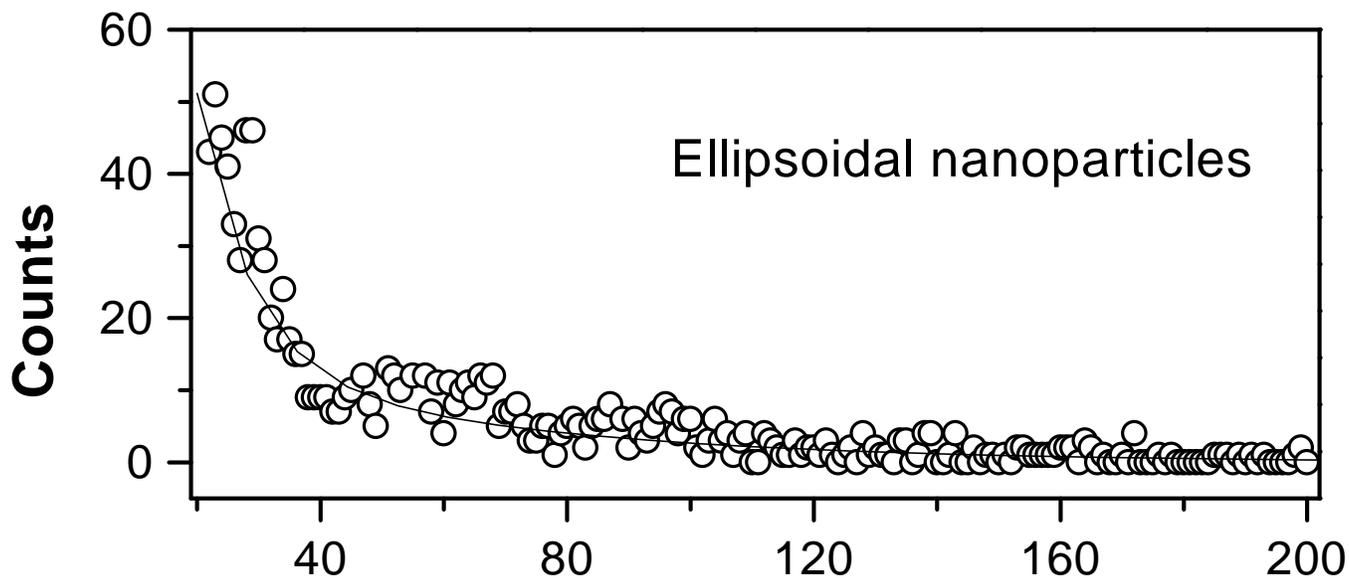
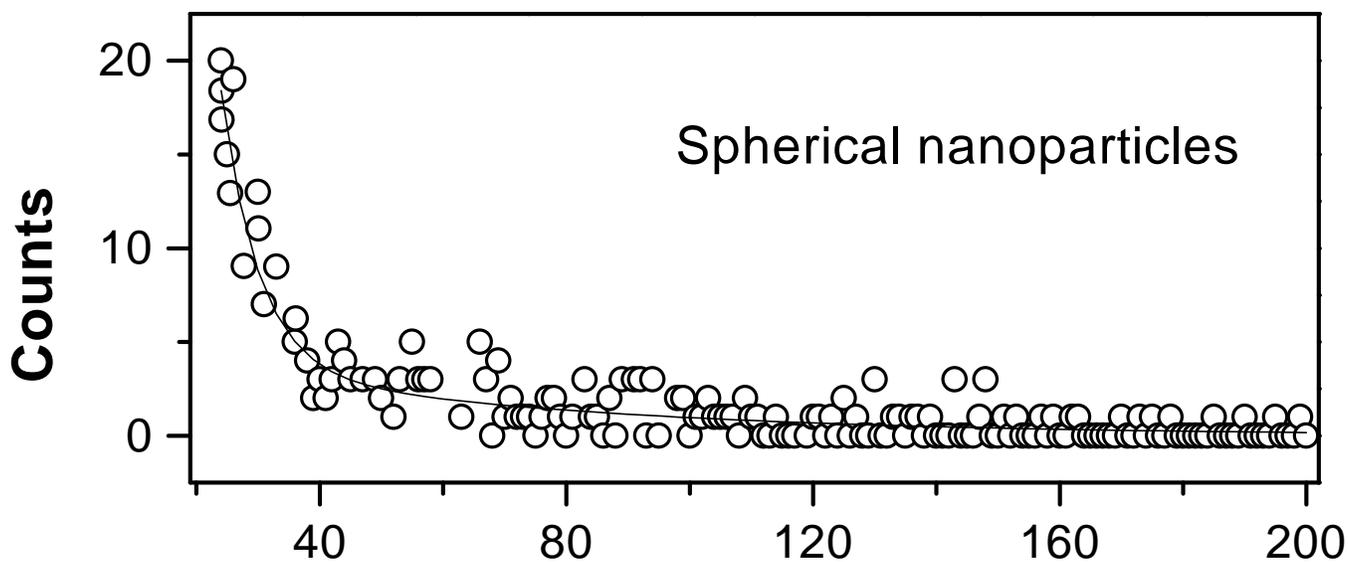
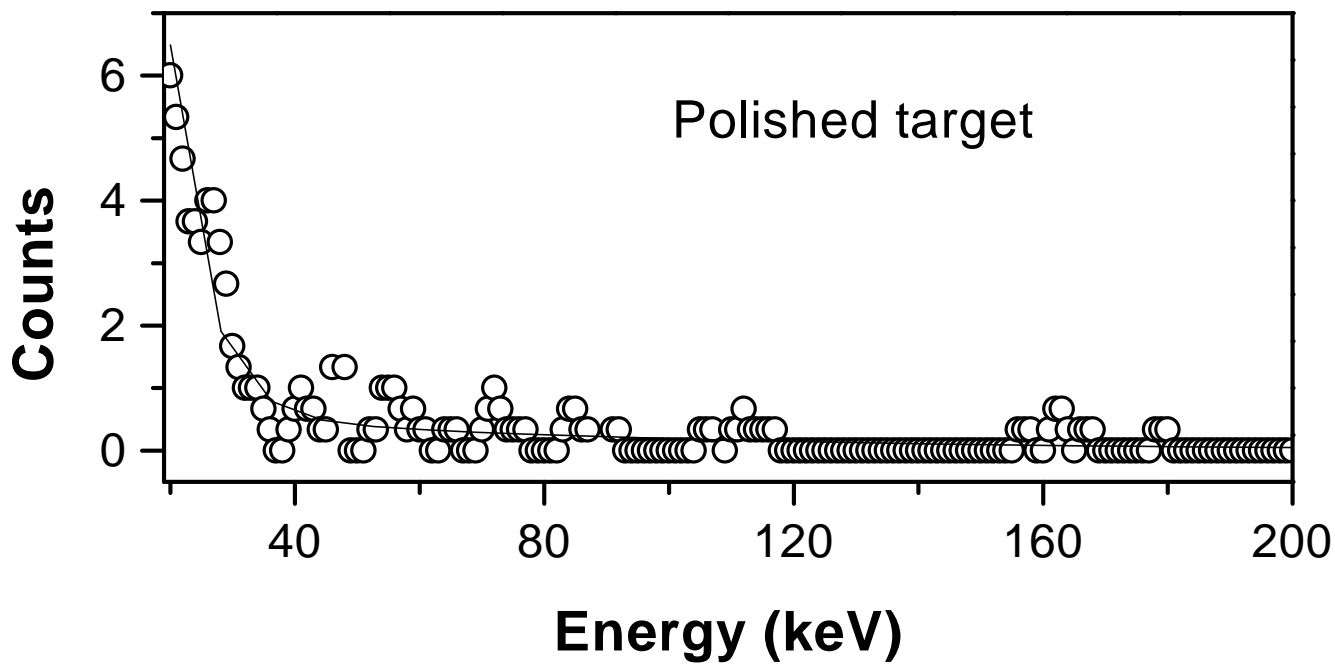
\end{document}